\documentclass[12pt]{iopart}

\newcommand{\bfm}[1]{\mbox{\boldmath${#1}$}}
\newcommand{\sect}[1]{\setcounter{equation}{0}\section{#1}}
\renewcommand{\theequation}{\arabic{section}.\arabic{equation}}
\newcommand{\app}{\setcounter{section}{0}
\setcounter{equation}{0} \renewcommand{\thesection}{APPENDIX
\Alph{section}}\renewcommand{\theequation}{\Alph{section}.
\arabic{equation}}}
\newcommand{\subsect}[1]{\setcounter{equation}{0}\subsection{#1}
\renewcommand{\theequation}{\arabic{section}.\arabic{subsection}.\arabic{equation}}}
\usepackage{graphicx}
\usepackage{dcolumn}

\begin{document}

\title[Basic-deformed thermostatistics]{Basic-deformed thermostatistics}
\author{A. Lavagno$^{1,2}$, A.M. Scarfone$^{1,3}$, and P. Narayana Swamy$^4$}
\address{$^1$Dipartimento di Fisica, Politecnico di Torino, Italy.}
\address{$^2$Istituto Nazionale di Fisica Nucleare (INFN),
Sezione di Torino, Italy.} \address{$^3$Istituto Nazionale di Fisica
della Materia (CNR--INFM), Sezione di Torino, Italy.}
\address{$^4$Department of Physics, Southern Illinois
University,
 Edwardsville, IL 62026, U.S.A.}

\date{\today}

\begin {abstract}
Starting from the {\em basic}-exponential, a $q$-deformed version of
the exponential function established in the framework of the {\em
basic}-hypergeometric series, we present a possible formulation of a
generalized statistical mechanics. In a $q$-nonuniform lattice we
introduce the {\em basic}-entropy related to the {\em
basic}-exponential by means of a $q$-variational principle.
Remarkably, this distribution exhibits a {\it natural} cut-off in
the energy spectrum. This fact, already encountered in other
formulations of generalized statistical mechanics, is expected to be
relevant to the applications of the theory to those systems governed
by long-range interactions. By employing the $q$-calculus, it is
shown that the standard thermodynamic functional relationships are
preserved, mimicking, in this way, the mathematical structure of the
ordinary thermostatistics which is recovered in the $q\to1$ limit.
\end {abstract}

\submitto{JPA} 

\eads{\mailto{andrea.lavagno@polito.it, antonio.scarfone@polito.it,
pswamy@siue.edu}} \maketitle


\sect{Introduction}

There has recently been a great deal of interest in investigating
deformed thermodynamic systems at the classical level. Such deformed
theories are believed to deal with the statistical behavior of
complex systems, whose underlying dynamics is spanned in a
multi-fractal phase-space, governed by long-range interaction and/or
long-time memory effects \cite{Beck,Hilborn,Abe01,GellMann}. A
possible fruitful mechanism capable of generating a deformed version
of the classical statistical mechanics consists of replacing, in the
Boltzmann-Gibbs distribution, the standard exponential with its
deformed version, accordingly postulating a deformed entropic form
which implies a generalized thermostatistics theory. In this manner,
some noteworthy generalizations of the standard statistical
mechanics have been proposed
\cite{Tsallis1,Abe,Kaniadakis1,Scarfone1} and their physical
consequences are currently under investigation
\cite{Scarfone03,Scarfone2,Tsallis2}.

On the other hand, quantum algebra and quantum groups have been the
subject of intensive research in several physical fields from cosmic
strings and black holes to the fractional quantum Hall effect and
high-$T_c$ superconductors \cite{wil}. From the seminal work of
Biedenharn \cite{bie} and Macfarlane \cite{mac}, it was clear that
the $q$-calculus, originally introduced by Heine \cite{heine} and by
Jackson \cite{jack} in the study of the basic hypergeometric series
\cite{gasper,Exton}, plays a central role in the representation of
the quantum groups with a deep physical meaning and not merely a
mathematical exercise \cite{cele1,fink}. In this context, it was
shown in Ref.\cite{Lavagno1}, that a natural realization of the
thermostatistics of $q$-deformed bosons and fermions can be built on
the formalism of $q$-calculus.

Furthermore, it is remarkable to observe that the $q$-calculus is
very well suited for to describe fractal and multifractal systems.
As soon as the system exhibits a discrete-scale invariance, the
natural tool is provided by Jackson $q$-derivative and $q$-integral,
which constitute the natural generalization of the regular
derivative and integral for discretely self-similar systems
\cite{erzan1}. In fact, it was shown that $q$-integral is related to
the free energy of spin systems on a hierarchical lattice
\cite{erzan2}.

In the recent past, some tentative to constructing a classical
counterpart to the quantum group and $q$-deformed dynamics have been
investigated \cite{marmo1}. Most recently, in
 \cite{noi1}, a $q$-deformed Poisson bracket has been
developed whose underlying algebra arises from the quantum group
theory appears to be invariant under the action of the
$q$-symplectic group. This generalization implies a classical
deformed dynamics and a deformed Fokker-Planck equation \cite{noi3}
whose stationary solution can be expressed in terms of the {\em
basic}-exponential, the $q$-analogue of the exponential function in
the framework of the basic hypergeometric series (henceforward, we
will use the term {\em basic}-exponential and {\em
basic}-thermostatistics to avoid any confusion with the
$q$-exponential customarily employed in the Tsallis'
thermostatistics formulation \cite{Tsallis1}).

The previous investigations raise the interesting question whether
the $q$-calculus and the {\em basic}-exponential can be introduced,
as a starting point, for the study of a deformed statistical
thermodynamics at the classical level. Just as the quantum
$q$-deformation plays a crucial role in the interpretation of
several complex physical systems, we expect that a classical
$q$-deformation of the thermostatistics can be relevant in several
physical applications. It is thus worthwhile to investigate the
structure of a classical statistical mechanics where the probability
distribution function is given by employing the {\em
basic}-exponential. This investigation represents the primary goal
of this paper.

Relating to the previously quoted generalized statistical theories
existing in the literature
\cite{Tsallis1,Abe,Kaniadakis1,Scarfone1}, it must be stressed that
the deformed exponential functions are very different from the one
we are introducing in the present work. The difference arises mainly
in the asymptotic behavior of the {\em basic}-exponential which
fails to show a power-law tail. Nevertheless, some mathematical
peculiarities exhibited by the {\em basic}-exponential makes it
relevant in the construction of a generalized statistical mechanics.
First among them, the existence of a {\em natural} cut-off in the
energy (velocity) spectrum which appears to be closely related to
the presence of long-range interactions among constituents of the
system. To the best of our knowledge, this was first suggested in
Ref.\cite{Spitzer} in order to overcome the infinity arising in the
mass and radius of an isothermal globular cluster. More recently,
concerning several physical systems, there have been many
investigations
\cite{boer,Plastino2,lee,Lavagno0,kami,vavro,naza,loeb} on the
relevance of the presence of an energy cut-off behavior in the
particle distribution functions.

Finally, we should stress that the {\em basic}-exponential
introduced in the present work is also employed in the formulation
of other interesting generalizations of physical theories such as,
for instance, in the $q$-deformed Schr\"odinger theory describing
the deformed quantum harmonic oscillator \cite{Wess} or in the
$q$-deformed theory of quantum coherence of bosons
\cite{Kulish,Yang}.

Our paper is organized as follows. In Section II, for convenience of
the reader, we review the main mathematical properties concerning
the quantum algebra of real numbers. We introduce the {\em
basic}-exponential by means of a power series and derive its main
proprieties which will be used in the formulation of the theory. In
Section III, we introduce the {\em basic}-entropy and, by means of a
$q$-version of the variational principle, we determine the deformed
distribution, for each case in turn, of a microcanonical, canonical
and grand canonical system. The thermodynamic structure of the
theory is explored in Section IV, while in Section V, we apply the
present formalism to a system of $q$-interacting particles whose
$q\to1$ limit reduces to the free system. Conclusions are reported
in Section VI and we end the paper with two mathematical Appendices.


\sect{Mathematical background}

We shall begin by recalling the main features of the $q$-calculus
for real numbers. It is based on the following $q$-commutative
relation among the operators $\hat x$ and $\hat\partial_x$,
\begin{eqnarray}
\hat\partial_x\hat x=1+q\,\hat x\,\hat\partial_x \ ,\label{lei4}
\end{eqnarray}
with $q$ a real and positive parameter.\\ A realization of the above
algebra in terms of ordinary real numbers can be accomplished by the
replacement \cite{Ubriaco}
\begin{eqnarray}
&&\hat x\to x \ ,\label{rap1}\\
&&\hat\partial_x\to{\cal D}_x \ ,\label{rap2}
\end{eqnarray}
where ${\cal D}_x$ is the Jackson derivative \cite{jack} defined as
\begin{equation}
{\cal D}_x=\frac{q^{x\,\partial_x}-1}{(q-1)\,x} \ .
\end{equation}
Its action on an arbitrary real function $f(x)$ is given by
\begin{equation}
{\cal D}_x\,f(x)=\frac{f(q\,x)-f(x)}{(q-1)\,x} \ .
\end{equation}
The Jackson derivative satisfies some simple proprieties which will
be useful in the following. For instance, its action on a monomial
$f(x)=x^n$ is given by
\begin{equation}
{\cal D}_x\,x^n=[n]\,x^{n-1} \ ,\label{7}
\end{equation}
and
\begin{equation}
{\cal D}_x\,x^{-n}= -{[n]\over q^n}\,{1\over x^{n+1}} \
,\label{7bis}
\end{equation}
where $n\geq0$ and
\begin{equation}
[n]=\frac{q^n-1}{q-1} \ ,\label{basic}
\end{equation}
are {\em basic}-numbers. By linearity, we can extend the action of
Jackson derivative to a generic polynomial. Moreover, we can easily
verify the following $q$-version of the Leibnitz rule
\begin{eqnarray}
\nonumber {\cal D}_x\Big(f(x)\,g(x)\Big)&=&{\cal
D}_x\,f(x)\,g(x)+f(q\,x)\,{\cal
D}_x\,g(x) \ ,\\
&=&{\cal D}_x\,f(x)\,g(q\,x)+f(x)\,{\cal D}_x\,g(x) \ .\label{leib}
\end{eqnarray}

A relevant role in the $q$-algebra, as developed by Jackson, is
given by the {\em basic}-binomial series defined by
\begin{eqnarray}
\nonumber
(x+y)^{(n)}&=&(x+y)\,(x+q\,y)\,(x+q^2\,y)\ldots(x+q^{n-1}\,y)\\
&\equiv&\sum_{r=0}^n\Big[^{_{\displaystyle n}}_{^{\displaystyle
r}}\Big]\,q^{r\,(r-1)/2}\,x^{n-r}\,y^r \ ,\label{qbin}
\end{eqnarray}
where
\begin{equation}
\Big[^{_{\displaystyle n}}_{^{\displaystyle
r}}\Big]=\frac{[n]!}{[r]!\,[n-r]!} \ ,\label{qbin1}
\end{equation}
is known as the $q$-binomial coefficient which reduces to the
ordinary binomial coefficient in the $q\to1$ limit \cite{Exton}. We
should remark that Eq.(\ref{qbin1}) holds for $0\leq r\leq n$, while
it is assumed to vanish otherwise and we have defined
$[n]!=[n]\,[n-1]\ldots[1]$. Equation (\ref{qbin}) can be easily
generalized to an arbitrary polynomial as shown in Appendix B.\\
Remarkably, a $q$-analogue of the Taylor expansion has been
introduced in Ref. \cite{jack} by means of a {\em basic}-binomial
(\ref{qbin}) as
\begin{eqnarray}
\hspace{-10mm}f(x)=f(a)+{(x-a)^{(1)}\over[1]!}\,{\cal
D}_x\,f(x)\Big|_{x=a}+{(x-a)^{(2)}\over[2]!}\,{\cal
D}_x^2\,f(x)\Big|_{x=a}+\ldots \ ,\label{Taylor}
\end{eqnarray}
where ${\cal D}_x^2\equiv{\cal D}_x\,{\cal D}_x$ and so on.\\
Consistently with the $q$-calculus, we also introduce the {\em
basic}-integration
\begin{equation}
\int\limits_0\limits^{\lambda_0}f(x)\,d_qx=\sum_{n=0}^\infty
\Delta_q \lambda_n\,f(\lambda_n) \ ,\label{qint1}
\end{equation}
where $\Delta_q \lambda_n=\lambda_n-\lambda_{n+1}$ and
$\lambda_n=\lambda_0\,q^n$ for $0<q<1$ whilst $\Delta_q
\lambda_n=\lambda_{n-1}-\lambda_n$ and
$\lambda_n=\lambda_0\,q^{-n-1}$ for $q>1$
\cite{gasper,Exton,erzan1}. Clearly, Eq.(\ref{qint1}) is reminiscent
of the Riemann quadrature formula performed now in a $q$-nonuniform
hierarchical lattice with a variable step $\Delta_q \lambda_n$. It
is trivial to verify that
\begin{equation}
{\cal D}_x\int\limits_0\limits^x f(y)\,d_qy=f(x) \ ,
\end{equation}
for any $q>0$.

Let us now introduce the following $q$-deformed function defined by
the series
\begin{equation}
{\rm E}_q(x)= \sum_{k=0}^{\infty}\, \frac{x^k}{[k]!}=1+x+
\frac{x^2}{[2]!}+\frac{x^3}{[3]!}+\cdots \ ,\label{qexp}
\end{equation}
which will play the main role in the framework we are introducing.
The function (\ref{qexp}) defines the {\em basic}-exponential, well
known in the literature since a long time ago, originally introduced
in the study of basic hypergeometric series
\cite{heine,gasper,Exton}. In this context, let us observe that
definition (\ref{qexp}) is fully consistent with
its Taylor expansion, as given by Eq. (\ref{Taylor}).\\
In the following, we briefly review the main algebraic properties of
the
{\it basic}-exponential useful for our developments.\\
The {\em basic}-exponential is a monotonically increasing function,
$d\,{\rm E}_q(x)/dx>0$, convex, $d^2{\rm E}_q(x)/dx^2>0$, with ${\rm
E}_q(0)=1$ and reducing to the ordinary exponential in the $q\to1$
limit: ${\rm
E}_1(x)\equiv \exp(x)$. \\
An important property satisfied by the $q$-exponential can be
written formally as \cite{Exton}
\begin{equation}
{\rm E}_q(x+y)={\rm E}_q(x)\,{\rm E}_{1/q}(y) \ ,\label{com1}
\end{equation}
where the left hand side of Eq.(\ref{com1}) must be considered by
means of its series expansion in terms of {\em basic}-binomials:
\begin{equation}
{\rm E}_q(x+y)=\sum_{k=0}^\infty{(x+y)^{(k)}\over[k]!} \
.\label{expq1}
\end{equation}
By observing that $(x-x)^{(k)}=0$ for any $k>0$, since
$(x-x)^{(0)}=1$, from Eq.(\ref{com1}) we obtain
\begin{equation}
{\rm E}_q(x)\,{\rm E}_{1/q}(-x)=1 \ .\label{inv}
\end{equation}

From the above relations, we easily deduce that, if $q<1$, the

series (\ref{qexp}) converges for all finite values of $x<1/(1-q)$,
otherwise, if $q>1$ the series
converges for $x>q/(1-q)$. \\
Thus, we can summarize the asymptotic behavior of the {\em
basic}-exponential as
\begin{eqnarray}
&&{\rm E}_q(-\infty)=0 \ ,\hspace{15mm}{\rm
E}_q\left({1\over1-q}\right)=+\infty \ ,\hspace{10mm}{\rm
if}\,\,q<1 \ ,\label{a1}\\
&&{\rm E}_q\left({q\over1-q}\right)=0 \ ,\hspace{10mm}{\rm
E}_q(+\infty)=+\infty \ ,\hspace{15mm}{\rm if}\,\,q>1 \ ,\label{a2}
\end{eqnarray}
where it is important to mention that the first expression of
Eq.(\ref{a2}) defines a cut-off condition of the {\em
basic}-exponential in the region relevant for our following
developments.

Among many properties, it is important to recall the following
relation \cite{Exton}
\begin{equation}
{\cal D}_x {\rm E}_q(a\,x)= a\,{\rm E}_q(a\,x) \ ,\label{JDE}
\end{equation}
and its dual
\begin{equation}
\int\limits_0\limits^x{\rm E}_q(a\,y)\,d_qy={1\over a}\Big[{\rm
E}_q(a\,x)-1\Big] \ .\label{dae}
\end{equation}
It must be pointed out that Eqs.(\ref{JDE}) and (\ref{dae}) are two
important properties of the {\em basic}-exponential which turns out
to be not true if we employ the ordinary derivative or integral. In
particular, from Eq.(\ref{JDE}) for $a=1$, we can obtain the further
useful relation
\begin{equation}
{\rm E}_q(q\,x)=\Big [1+(q-1)\,x\Big ]\,{\rm E}_q(x) \ .\label{JDE1}
\end{equation}
Moreover, from Eq.(\ref{dae}) with $a<0$, we have
\begin{equation}
\int\limits_0\limits^{x_{\rm max}}{\rm E}_q(a\,y)\,d_qy=-{1\over a}
\ ,
\end{equation}
where $x_{\rm max}\to+\infty$ in the $q<1$ case, while $x_{\rm
max}=q/(q-1)$ in the $q>1$ case, accounting for the cut-off
condition (\ref{a2}).

In addition to the {\em basic}-exponential, we can also introduce
 its inverse function, the {\em basic}-logarithm ${\rm Ln}_q(x)$,
 such that
\begin{equation}
{\rm E}_q\Big({\rm Ln}_q(x)\Big)={\rm Ln}_q\Big({\rm E}_q(x)\Big)=x
\ ,\label{qlog}
\end{equation}
which certainly exists because ${\rm E}_q(x)$ is a strictly
monotonic function. Many proprieties of the {\em basic}-logarithm
follow directly from the corresponding ones of the {\em
basic}-exponential. For instance, ${\rm Ln}_q(x)$ is a monotonic,
increasing and concave function ($d\,{\rm Ln}_q(x)/d\,x>0$,
$d^2\,{\rm Ln}_q(x)/d\,x^2<0$), normalized in ${\rm Ln}_q(1)=0$ and
the asymptotic behavior given by
\begin{eqnarray}
&&{\rm Ln}_q(0)=-\infty \ ,\hspace{10mm}{\rm
Ln}_q\left(+\infty\right)={1\over1-q} \ ,\hspace{10mm}{\rm
if}\,\,q<1 \ ,\\
&&{\rm Ln}_q\left(0\right)={q\over1-q} \ ,\hspace{7mm}{\rm
Ln}_q(+\infty)=+\infty \ ,\hspace{13mm}{\rm if}\,\,q>1 \ .
\end{eqnarray}
Although a definition of ${\rm Ln}_q(x)$ through a series is
possible, it appears to be a nontrivial task to write it in an easy
form and, to the best of our knowledge, there are no definitive
results in the literature (see for instance Ref.\cite{Charles}).

We conclude this Section by remarking that alternative definitions
of the {\em basic}-exponential, by means of a different definition
of {\em basic}-numbers, has been widely employed in the literature
\cite{Ward}. Among the many, we may point out the choice based on
the symmetric definition $[n]_{\rm s}=(q^n-q^{-n})/(q-q^{-1})$. The
corresponding symmetric {\em basic}-exponential may be defined on
the whole real region $(-\infty,\,+\infty)$ and has the symmetry
$q\to1/q$. As a consequence, the symmetric {\em basic}-exponential
does not present the cut-off feature.


\sect{Basic-entropy and its distribution}

Equipped with the {\em basic}-functions, our aim is to formulate a
statistical mechanics based on the formalism of the
$q$-algebra and to study its main physical implications.\\
On the basis of the above mathematical framework, it appears natural
to generalize the Boltzmann entropy to the following form
\begin{equation}
S_{q}(p)=-\int\limits_{\cal M} p(\lambda)\,{\rm
Ln}_q\,p(\lambda)\,d_q\lambda \ ,\label{qent}
\end{equation}
where $p(\lambda)$ is probability distribution function labeled by a
set of parameters $\lambda$ running on the manifold ${\cal M}$,
eventually identified with the phase space, which define the
accessible states of the system. Henceforward we adopt
units where the Boltzmann constant $k_{\rm B}=1$.\\
Equation (\ref{qent}) resembles the well-known Boltzmann-Gibbs
entropy $S^{\rm BG}(p)$ through the replacement of the logarithm
with the {\em basic}-logarithm. Clearly, the function $S_q(p)$
reduces to the standard entropy $S^{\rm BG}(p)\equiv S_1(p)$ in the
$q\to1$ limit. In the following, we shall refer to the function
(\ref{qent}) as {\em basic}-entropy.

A way to obtain the equilibrium distribution from the entropy
$S_q(p)$, given a set of $M+1$ constraints $\Phi_j(p)$ with
$j=0,\,\ldots,\,M$, can be accomplished through the following
variational problem
\begin{equation}
\delta{\mathcal F}(p)=0 \ ,\label{var}
\end{equation}
with
\begin{equation}
{\mathcal F}(p)=\left( S_{q}(p)-\sum_j\mu^\ast_j\,\Phi_j(p)\right) \
,\label{var2}
\end{equation}
where $\mu^\ast_j$ are the Lagrangian multipliers associated with
the constraints $\Phi_j(p)$.\\ Quite generally, such constraints can
be written in
\begin{equation}
\Phi_j(p)=\int\limits_{\cal M}
\phi_j(\lambda)\,p(\lambda)\,d_q\lambda \ ,\label{constr}
\end{equation}
representing the mean value of the quantities $\phi_j(\lambda)$
which are identifiable with suitably physical observable. In
particular, for $\phi_0(\lambda)=1$, Eq.(\ref{constr}) give the
normalization of the distribution function.\\ In order to handle
easily the variation problem of Eq.(\ref{var}), let us introduce the
ansatz
\begin{equation}
p(\lambda)={\rm E}_q\Big(-f(\lambda)\Big) \ .\label{ans}
\end{equation}
We recall that ${\rm E}_q(-x)$ is a strictly monotonically
decreasing function. This means that it reaches its minimum at
points which maximize the functions $f(\lambda)$.\\ Actually, the
problem Eq.(\ref{var}) can be replaced by the following equivalent
$q$-variational problem
\begin{equation}
\delta_q\widetilde{\mathcal F}(f)={\cal D}_f \, \widetilde{\mathcal
F}(f)\,\Delta_q f=0 \ ,\label{var1}
\end{equation}
with $\Delta_q f=(q-1) f$ and
\begin{equation}
\widetilde{\mathcal F}(f)\equiv {\mathcal
F}\big(p(f)\big)=\int\limits_{\mathcal M}{\rm
E}_q\Big(-f(\lambda)\Big)\,\Big(f(\lambda)-\sum_j\mu_j\,\phi_j(\lambda)\Big)
\,d_q\lambda \ .
\end{equation}
It is show in Appendix A that, according to the $q$-algebra, both
the calculus (\ref{var}) and (\ref{var1}) give substantially the
same result (apart a redefinition of the Lagrange multipliers,
$\mu_j\to\mu^\ast_j$, which, accounting for the relevant constraint
equations, has no any effect on the expression of the final
distribution). Equation (\ref{var1}) can solve more speedily as
\begin{eqnarray}
\nonumber \delta_q\widetilde{\mathcal F}(p)&&=\left[{\cal
D}_f\int\limits_{\cal M}{\rm
E}_q\Big(-f(\lambda)\Big)\,\Big(f(\lambda)-\sum_j\mu_j\,\phi_j(\lambda)\Big)\,d_q\lambda\right]\Delta_q f\\
&&=\left[\int\limits_{\cal M}{\rm
E}_q\Big(-f(\lambda)\Big)\,\Big(q\,f(\lambda)-1-\sum_j\mu_j\,\phi_j(\lambda)\Big)\,d_q\lambda\right]\Delta_qf=0
\ ,
\end{eqnarray}
which implies the following relation
\begin{equation}
f(\lambda)=q^{-1}\Big(1+\sum_j\mu_j\,\phi_j(\lambda)\Big) \
,\label{distr}
\end{equation}
and accounting for Eq. (\ref{ans}) we obtain the general solution in
the form
\begin{equation}
p(\lambda)={\rm
E}_q\Bigg(-q^{-1}\,\Big(1+\sum_j\mu_j\,\phi_j(\lambda)\Big)\Bigg) \
.
\end{equation}
It is easy to verify that this expression reduces, in the $q\to1$
limit, to the ordinary Gibbs distribution. Let us consider
separately three main cases: the microcanonical system, the
canonical system and the grand canonical system.

\subsect{Microcanonical system}

We consider a closed system with a given fixed energy $U$, volume
$V$ and particle number $N$. In this case the system is only forced
by the constraint
\begin{equation}
\int\limits_{\cal M} p(\lambda)\,d_q\lambda=1 \ ,\label{norm}
\end{equation}
which assures the normalization of the distribution.\\
Before to proceed on, let us spend a few word about Eq.(\ref{norm})
to better understand the underlying physical meaning of the
$q$-calculus. By taking into account the definition of the {\em
basic}-integral (\ref{qint1}), for $0<q<1$, we obtain
\begin{equation}
\int\limits_{\cal M}
p(\lambda)\,d_q\lambda=\lambda_0\,(1-q)\left[p(\lambda_0)+q\,p(\lambda_1)
+q^2\,p(\lambda_2)+\cdots\right] \ ,\label{norm1}
\end{equation}
where $\lambda_0$ is a constant and $\lambda_i=q^i\,\lambda_0$. This
expression shows two important features of the theory we are
developing. First, the $q$-integral plunges, in a natural way, to
consider a $q$-deformed lattice whose amplitude of the elementary
cell $\Delta_q\lambda_0=\lambda_0\,(1-q)$ is shrunken, step by step,
by the quantity $q^n$. This is a substantial different situation
with respect to the standard case obtained in the $q\to1$ limit
where any probability (differential probability) is multiplied by an
equal (infinitesimal) quantities: $dp(x)=p(x)\,dx$. Second, the
parameter space assumes a fractal structure given by the rule
$\lambda_n=q^n\,\lambda_0$. We recognize a self-similarity in the
parameter space since, starting from any level $N>1$ the same
structure
$\lambda_{N+n}=q^n\,\lambda_N$ is discovered.\\
Remarkably, by introducing a set of discrete probability
distributions $p_n$, related to $p(\lambda)$, as
\begin{equation}
p_n=\Delta_q\lambda_n\,p(\lambda_n) \ ,\label{discr1}
\end{equation}
equation (\ref{norm}) becomes
\begin{equation}
\sum_{i=0}^\infty p_i=1 \ .\label{discr}
\end{equation}
Clearly, the same considerations hold also in the $q>1$ case.\\ In
order to derive the expression for $p(\lambda)$ we introduce the
following constrained entropic functional
\begin{equation}
{\cal F}(p)=-\int\limits_{\cal M} p(\lambda)\Big[{\rm
Ln}_q\,p(\lambda)+\alpha\Big]\,d_q\lambda \ ,\label{fmicro}
\end{equation}
where $\alpha$ is the Lagrange multiplier associated with Eq.(\ref{norm}).\\
By inserting the ansatz (\ref{ans}) in Eq.(\ref{fmicro}) we obtain
\begin{equation}
\widetilde{\cal F}(f)=\int\limits_{\cal M} {\rm
E}_q\Big(-f(\lambda)\Big)\,\Big(f(\lambda)-\alpha\Big)\,d_q\lambda \
.
\end{equation}
By evaluating the equation $\delta_q\,\widetilde{\cal F}(f)=0$,
accounting for Eqs.(\ref{leib}) and (\ref{JDE}), we obtain
\begin{equation}
\delta_q\widetilde{\cal F}(f)={\rm
E}_q\Big(-f(\lambda)\Big)\,\Big(1+\alpha-q\,f(\lambda)\Big)=0 \
.\label{qme1}
\end{equation}
The above expression implies the following equation
\begin{equation}
{\rm Ln}_q\,p(\lambda)+q^{-1}\,(1+\alpha)=0 \ ,\label{qme2}
\end{equation}
which gives the microcanonical distribution in the form
\begin{equation}
p(\lambda)={\rm E}_q\left(-q^{-1}\,(1+\alpha)\right) \ .\label{mm}
\end{equation}
[We refer to Appendix B for a systematic derivation of
Eq.(\ref{mm})].\\
Since the above expression does not depend on $\lambda$, by
accounting for the condition (\ref{norm}), we obtain the expected
microcanonical uniform distribution
\begin{equation}
p={1\over W} \ ,
\end{equation}
where the number of accessible states
\begin{equation}
W(U,\,V,\,N)=\int\limits_{\cal M} d_q\lambda \ ,
\end{equation}
is related to the Lagrange multiplier through the relation
$\alpha=-1-q\,{\rm Ln}_q(1/W)$, which is a function of the energy,
the volume and the total number of particles in the system.


\subsect{Canonical system}

By following the steps described in the previous Subsection, we can
derive the canonical distribution for an open system that exchanges
energy with the surrounding. In this case, let us pose
$\epsilon\equiv\epsilon(\lambda)$ the energy of the system still
labeled through $\lambda$ so that, among to the constraint
(\ref{norm}), we impose the further condition on the mean energy
\begin{equation}
\int\limits_{\mathcal M}
\epsilon(\lambda)\,p\big(\epsilon(\lambda)\big)\,d_q\lambda=\langle\epsilon\rangle
\ .\label{energy}
\end{equation}
Accounting for the definition (\ref{qint1}) this last condition
becomes ($0<q<1$)
\begin{equation}
\lambda_0\,(1-q)\,\left[\epsilon_0\,
p(\epsilon_0)+q\,\epsilon_1\,p(\epsilon_1)+q^2\,\epsilon_2\,p(\epsilon_2)+\cdots\right]=\langle\epsilon\rangle
\ ,
\end{equation}
where $\epsilon_i\equiv\epsilon(\lambda_i)$ so that the $q$-calculus
implies a fractal structure in the energy spectrum. With the
position (\ref{discr1}) we realize that the constraint
(\ref{energy}) on the mean energy is equivalent to the standard
definition
\begin{equation}
\sum_{i=0}^\infty\epsilon_i\,p_i=\langle\epsilon\rangle \ ,
\end{equation}
although the fractal structure in the energy spectrum still holds
being implicitly contained
in the definition of the discrete probabilities $p_i$.\\
 After introducing the constrained entropic functional
\begin{equation}
{\cal F}(p)=-\int\limits_{\mathcal M}
p\big(\epsilon(\lambda)\big)\Big[{\rm
Ln}_q\,p\big(\epsilon(\lambda)\big)+\alpha+\beta\,\epsilon(\lambda)\Big]\,d_q\lambda
\ ,\label{fcan}
\end{equation}
with $\alpha$ and $\beta$ the Lagrange multiplier associated with
Eqs.(\ref{norm}) and (\ref{energy}), by imposing the ansatz
(\ref{ans}) we are dealing with the variational problem
$\delta_q\widetilde{\cal F}(f)=0$ where
\begin{equation}
\widetilde{\cal F}(f)=\int\limits_{\mathcal M} {\rm
E}_q\Big(-f\big(\epsilon(\lambda)\big)\Big)
\Big[f\big(\epsilon(\lambda)\big)-\alpha-\beta\,\epsilon(\lambda)\Big]\,d_q\lambda
\ ,
\end{equation}
and whose solution reads
\begin{equation}
{\rm
Ln}_q\,p(\epsilon)+q^{-1}\,\left(1+\alpha+\beta\,\epsilon\right)=0 \
.\label{eq1}
\end{equation}
According to the $q$-algebra described in Eq.(\ref{aa3}), with
$x={\rm Ln}_q\,p(\epsilon)$, $y=q^{-1}\,(1+\alpha)$ and
$z=q^{-1}\,\beta\,\epsilon$, from Eq.(\ref{eq1}) we derive the
following canonical distribution
\begin{equation}
p(\epsilon)={\rm E}_q\left(-q^{-1}\,(1+\alpha)\right)\,{\rm
E}_q\Big(-\beta_q\,\epsilon\Big) \ ,\label{can}
\end{equation}
where $\beta_q=q^{-1}\,\beta$ (see Appendix B). By imposing the
normalization condition (\ref{norm}) on the distribution
$p(\epsilon)$, we obtain
\begin{equation}
p(\epsilon)={1\over Z_{q}}\,{\rm E}_{q}\Big(-\beta_q\,\epsilon\Big)
\ ,\label{cdis}
\end{equation}
where $Z_{q}$ is the canonical partition function defined by
\begin{equation}
Z_{q}={\rm
E}_{1/q}\left(q^{-1}\,(1+\alpha)\right)=\int\limits_{\mathcal M}
{\rm E}_{q}\Big(-\beta_q\,\epsilon(\lambda)\Big)\,d_q\lambda \
.\label{cZ}
\end{equation}
Trivially, Eq.(\ref{cdis}) reduces to the
canonical Gibbs distribution in the $q\to1$ limit. \\
We remark that, for $q<1$ all the energy levels $\epsilon$ can be
occupied. For $q>1$, however, the distribution (\ref{cdis}) shows a
cut-off in the energy spectrum due to the finite convergence radius
of the function ${\rm E}_q(x)$. This is an important consequence of
the theory under investigation which limits the number of states
accessible to the system. Its origin can be related, as generally
accepted, to the presence of interactions among the parts of the
system, whose effect is to reduce the volume of the phase space.
This peculiarity is encountered also in other statistics mechanical
models based on generalized entropic forms.\\ In particular, Eq.
(\ref{a2}) imposes the following limiting condition on the energy
levels $\epsilon<\epsilon_{\rm max}$, where
\begin{equation}
\epsilon_{\rm max}=\frac{q}{(q-1)\,\beta_q} \ .\label{limit}
\end{equation}
Physically, this means that all the microscopic configurations of
the phase space corresponding to an energy $\epsilon$ beyond
$\epsilon_{\rm max}$ are statistically unattainable. We remark, that
$\epsilon_{\rm max}$ is a function of $\beta_q$ which plays the role
of the inverse of a pseudo temperature. Its value is determined
through Eq.(\ref{energy}) and it is expected, like in the undeformed
theory, that $\beta_q$ decreases as the mean energy increases. It
means that the cut-off condition plays a relevant role in those {\em
small} systems whose mean energy is small compared with the typical
energy values of the macroscopic systems. This is reminiscent of a
quantum scenario although our system is a classical one.
\begin{figure}[ht]
\begin{center}
   \resizebox{.8\textwidth}{!}{\includegraphics{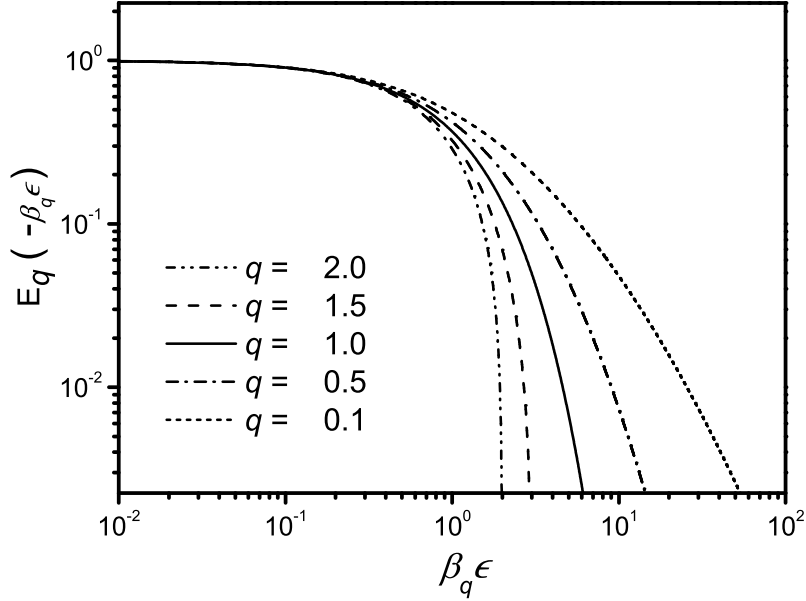}}
\caption{Generalized Boltzmann distribution ${\rm
E}_q(-\beta_q\,\epsilon)$ for different values of the deformed
parameter $q$. The solid line coincides with the standard
exponential function $\exp(-\beta\,\epsilon)$.}
  \end{center}
\end{figure}
In Figure 1, we illustrate the behavior of the basic Boltzmann
factor ${\rm E}_q(-\beta_q\,\epsilon)$ for different values of the
deformation parameter $q$ compared with the classical one given by
$q=1$. It is observed that for $q<1$, high energy events are
enhanced with respect to the standard case while, for $q>1$, events
are more and more inhibited when energy increases until it reaches
the cut-off
point, where $p(\epsilon)=0$.\\
\begin{figure}[ht]
\begin{center}
   \resizebox{.8\textwidth}{!}{\includegraphics{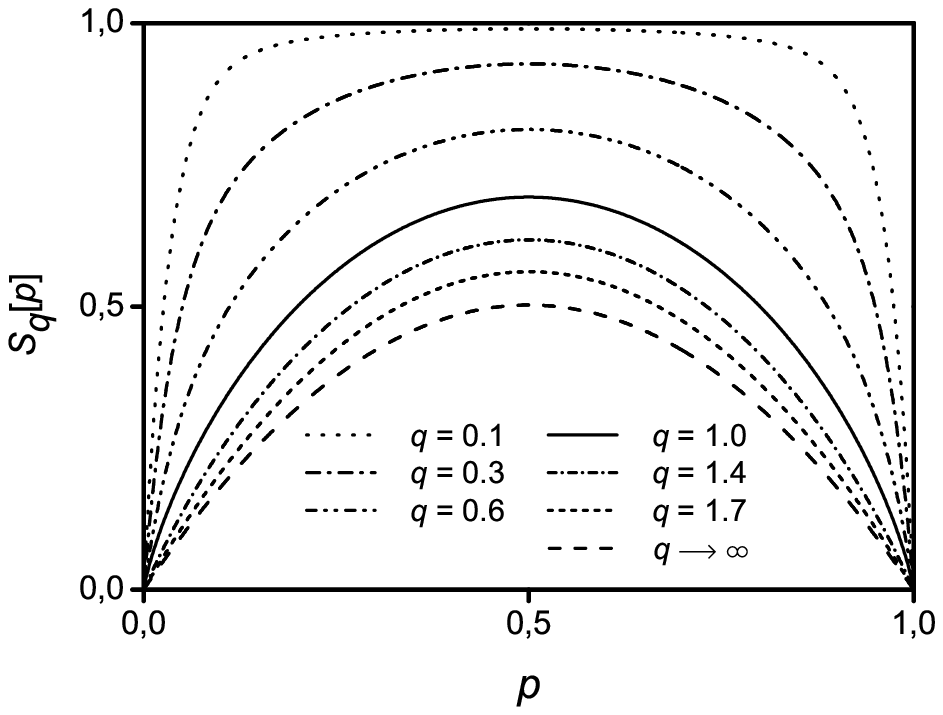}}
\caption{Plot of {\em basic}-entropy, for different values of the
deformation parameter, for a system with two levels. The dashed line
represents the asymptotic curve reached for $q\gg 1$.}
   \end{center}
\end{figure}
In Figure 2, we show the {\em basic}-entropy for a system of two
levels with probabilities $p$ and $1-p$, respectively, for different
values of the deformation parameter. The dashed line represents the
asymptotic curve reached for $q\gg 1$.\\
Again, we stress that such a feature is also shown by other
distributions obtained from physically motivated generalizations of
the Boltzmann-Gibbs entropy. For instance, it can be shown
\cite{Scarfone2} that for suitably chosen deformation parameters,
generalized entropies belonging to the two-parameter family of
Sharma-Mittal
 \cite{Sharma}, which include among the others also the
well known Tsallis' entropy \cite{Tsallis1} and the R\`{e}nyi's
entropy \cite{Renyi}, generate probability distribution functions
which exhibit a cut-off in their tails. Nevertheless, it is
worthwhile to observe that almost all the members belonging to the
Sharma-Mittal family, with some exception like the Boltzmann-Gibbs
entropy and the Gaussian entropy, have an asymptotic power law
behavior. This differs substantially from the asymptotic behavior
shown by {\em basic}-distribution, which is more similar to that of
the stretched exponential. In this respect, the theory under
investigation is not an alternative but complementary to the already
existing generalized version of the statistical mechanics, since it
can be relevant in the study of those complex systems which are not
characterized by an asymptotic free-scale behavior.


\subsect{Grand canonical system}

Finally, let us investigate the grand canonical distribution
describing an open system where energy and particles can be
exchanged with the surrounding. This can be accomplished by imposing
the following constraints
\begin{eqnarray}
&&\sum_N\int\limits_{\mathcal M}p_{_N}\big(\epsilon(\lambda)\big)\,d_q\lambda=1 \ ,\label{norm2}\\
&&\sum_N\int\limits_{\mathcal M}\epsilon(\lambda)\,p_{_N}
\big(\epsilon(\lambda)\big)\,d_q\lambda=\langle\epsilon\rangle \ ,\\
&&\sum_N \int\limits_{\mathcal
M}N\,p_{_N}\big(\epsilon(\lambda)\big)\,d_q\lambda=\langle N\rangle
\ ,\label{particle}
\end{eqnarray}
on the the normalization, the mean energy and the mean particle
number, where $N=1,\,\ldots,\,\infty$ enumerate the particles contained in the system.\\
The distribution function is found just as in the former cases. We
form the constrained entropy
\begin{equation}
{\cal F}(p)=-\sum_N\int\limits_{\mathcal
M}p_{_N}\big(\epsilon(\lambda)\big)\Big[{\rm
Ln}_q\,p_{_N}\big(\epsilon(\lambda)\big)+\alpha+\beta\,\epsilon(\lambda)+\gamma\,N\Big]\,d_q\lambda
\ ,\label{gg1}
\end{equation}
where $\alpha$, $\beta$ and $\gamma$ are the Lagrange multipliers
associated with the constraints (\ref{norm2})-(\ref{particle}),
respectively. By evaluating the equation $\delta_q\,\widetilde{\cal
F}(f)=0$, with
\begin{equation}
\widetilde{\cal F}(f)=\sum_N\int\limits_{\mathcal M}{\rm
E}_q\Big(-f_{_N}\big(\epsilon(\lambda)\big)\Big)\,
\Big[f_{_N}\big(\epsilon(\lambda)\big)-\alpha-\beta\,\epsilon(\lambda)
-\gamma\,N\Big]\,d_q\lambda \ ,
\end{equation}
obtained from Eq.(\ref{gg1}) by using the ansatz
$p_{_N}(\epsilon)={\rm E}_q\Big(-f_{_N}(\epsilon)\Big)$, we derive
the following result
\begin{equation}
{\rm
Ln}_q\,p_{_N}(\epsilon)+q^{-1}\,\left(1+\alpha+\beta\,\epsilon-\mu\,\beta\,N\right)=0
\ ,\label{gce}
\end{equation}
where we set $\gamma=-\mu\,\beta$. According to the $q$-algebra
(\ref{aa3}) with $x={\rm Ln}_q\,p_{_N}(\epsilon)$,
$y=q^{-1}\,(1+\alpha)$, $z=q^{-1}\,\beta\,\epsilon$ and
$u=-q^{-1}\,\mu\,\beta\,N$, from Eq.(\ref{gce}) we obtain the grand
canonical distribution function in the form
\begin{equation}
p_{_N}(\epsilon) = {1\over {\cal Z}_q}\,{\rm E}_q
\Big(-\beta_q\,\epsilon\Big)\,{\rm E}_q\Big(\mu\,\beta_q\,N\Big) \
,\label{gcan}
\end{equation}
where
\begin{equation}
{\cal Z}_{q}={\rm
E}_{1/q}\left(q^{-1}(1+\alpha)\right)=\sum_N\int\limits_{\mathcal M}
{\rm E}_q \Big(-\beta_q\,\epsilon(\lambda)\Big)\,{\rm
E}_q\Big(\mu\,\beta_q\,N\Big)\,d_q\lambda \ ,\label{gZ}
\end{equation}
is the grand partition function.


\sect{Some thermodynamic relations}
\renewcommand{\theequation}{\arabic{section}.\arabic{equation}}

In the following we shall investigate the thermodynamic structure of
the theory. It is shown that some basic relations of the standard
theory can be transcribed in a straightforward manner in the present
formalism establishing, in this way, the ground for a
generalized classical thermodynamics based on the framework of the $q$-algebra.\\
To begin with, we observe that by multiplying Eq.(\ref{gce}) by
$p_{_N}(\epsilon)$ and taking into account all the constraints
imposed on the system, we obtain the relation
\begin{equation}
S_{q}(\langle{\epsilon}\rangle,\,\langle N \rangle,\,V)= {\rm
Ln}_{1/q}\,{\cal
Z}_{q}+\beta_q\,\langle{\epsilon}\rangle-\mu\,\beta_q\,\langle
N\rangle \ ,\label{ent}
\end{equation}
where we set ${\rm Ln}_{1/q}\,{\cal Z}_{q}=q^{-1}\,(1+\alpha)$ as it
follows from Eq.(\ref{gZ}). Equation (\ref{ent}) mimics the standard
relation $S=\ln\,{\cal
Z}+\beta\,U-\mu\,\beta\,N$ which is recovered in the $q\to1$ limit.\\
If one is willing to identify the quantity $\beta_q$ with the
inverse of the temperature $\beta_q=1/T$, Eq.(\ref{ent}) can be
rewritten in
\begin{equation}
\langle{\epsilon}\rangle=T\,S_{q}-T\,{\rm Ln}_{1/q}\,{\cal
Z}_{q}\,+\mu\,\langle N\rangle \ .\label{us}
\end{equation}
On the other hand, in analogy with standard thermodynamics, we can
introduce the pressure $P$ and the volume $V$ of the system by
requiring that all the thermodynamics variables are functionally
related through the following relationship
\begin{equation}
\langle{\epsilon}\rangle=T\,S_{q}-P\,V+\mu\,\langle N\rangle \
,\label{euler}
\end{equation}
which is a constitutive equation for the theory under investigation
and can be identified with the $q$-analogue of the Euler's equation
\cite{Callen}. By comparing Eqs.(\ref{us}) and (\ref{euler}) we are
thus encouraged to define the pressure through the relation
\begin{equation}
P\,V=T\,{\rm Ln}_{1/q}\,{\cal Z}_{q} \ ,\label{state}
\end{equation}
which represents the $q$-generalized state equation for a system
described by the {\em basic}-entropy.

The partition function (\ref{cZ}), as well as the grand partition
function (\ref{gZ}), are useful tools to evaluate the statistical
proprieties of the system. In fact, by evaluating the Jackson
derivative of $Z_q$ as
\begin{equation}
{\cal D}_{\beta_q}\,Z_{q}={\cal D}_{\beta_q}\,\int\limits_{\mathcal
M}{\rm
E}_{q}\Big(-\beta_q\,\epsilon(\lambda)\Big)\,d_q\lambda=-\int\limits_{\mathcal
M}\epsilon(\lambda)\,{\rm
E}_{q}\Big(-\beta_q\,\epsilon(\lambda)\Big)\,d_q\lambda \ ,
\end{equation}
where Eq.(\ref{JDE}) has been used, we obtain the result
\begin{equation}
\langle\epsilon\rangle=-{1\over Z_{q}}\,{\cal D}_{\beta_q}\,Z_{q} \
,
\end{equation}
which, in the $q\to1$ limit reduces to the well-known relation
$\langle\epsilon\rangle=-(d Z/d\,\beta)/Z$.\\
Similar results can be obtained starting from the function ${\cal
Z}_{q}$ as follows
\begin{eqnarray}
&&\langle\epsilon\rangle=-{1\over {\cal Z}_{q}}\,{\cal D}_{\beta_q}\,{\cal Z}_{q} \ ,\\
&&\langle N\rangle={1\over {\beta_q\,\cal Z}_{q}}\,{\cal
D}_\mu\,{\cal Z}_{q} \ .
\end{eqnarray}
Incidently, by using the prescription
\begin{equation}
\epsilon \to\epsilon+\delta\,A(\epsilon) \ ,
\end{equation}
where $A(\epsilon)$ is an arbitrary physical observable, the
expectation value of $A$ can be obtained in
\begin{equation}
\langle A\rangle=-{1\over \beta_q\,Z_{q}}{\cal
D}_{\delta}\,Z_{q}\Bigg|_{\delta=0} \ .\label{meanvalue}
\end{equation}
In this sense the partition function encode all the statistical
information contained in the system.\\
Another physically relevant quantity is obtained starting from the
expression
\begin{equation}
{\cal D}_{\beta_q}\langle{\epsilon}\rangle={\cal
D}_{\beta_q}\left({1\over Z_{q}(\beta_q)}\int\limits_{\mathcal
M}\epsilon(\lambda)\,{\rm
E}_{q}\Big(-\beta_q\,\epsilon(\lambda)\Big)\,d_q\lambda\right) \ ,
\end{equation}
where we have explicitly indicated the dependence on $\beta_q$ in
the partition function. By taking into account Eq.(\ref{JDE1}), we
can derive the result
\begin{equation}
{\cal D}_{\beta_q}\langle
{\epsilon}\rangle=-\frac{\langle\epsilon^2\rangle-\langle
\epsilon\rangle^2}{1-(q-1)\,\beta_q\,\langle\epsilon\rangle}\equiv-\sigma^2_{q,\varepsilon}
\ ,\label{fluc}
\end{equation}
where $\langle\epsilon^2\rangle=\int \epsilon^2\,p(\epsilon)\,d_q\lambda$.\\
In the
$q\to1$ limit, we obtain the classical relation $d\,\langle\epsilon\rangle/d\,\beta=-\sigma^2_\varepsilon$.\\
The quantity $\sigma_{q,\varepsilon}$ measures the fluctuation of
the energy of the system around its mean value
$\langle\epsilon\rangle$. It is observed that, compared to the
classical case, such
fluctuations are reduced for $q<1$ and are enlarged for $q>1$.\\
Alternatively, by mimicking the classical definition, one can
introduce the heat capacity by
\begin{equation}
C_V=-\beta_q^2\,{\cal D}_{\beta_q}\langle\epsilon\rangle\equiv
(\beta_q\,\sigma_{q,\varepsilon})^2 \ ,
\end{equation}
so that, the relative width of the fluctuations in energy is given
by
\begin{equation}
\frac{\sigma_{q,\varepsilon}}{\langle\epsilon\rangle}={1\over
\beta_q\,\langle\epsilon \rangle}\sqrt{C_V} \ .
\end{equation}
Similar derivations can also be made for the grand canonical case
where, together with Eq.(\ref{fluc}) we can obtain also the further
relation
\begin{equation}
{\cal D}_{\mu}\langle N\rangle=\beta_q\,\frac{\langle
N^2\rangle-\langle N\rangle^2}{1+(q-1)\,\mu\,\beta_q\,\langle
N\rangle}\equiv\sigma^2_{q,N} \ ,\label{flucn}
\end{equation}
with $\langle N^2\rangle=\sum_{N}\int
N^2\,p_{_N}(\epsilon)\,d_q\lambda$, which measures
the deviation from the mean value $\langle N\rangle$.\\


\sect{Basic-ideal gas}

In this Section, to illustrate the consequences of our generalized
thermodynamical model, we discuss a simple {\em
basic}-noninteracting particle gas which reduces, in the $q\to1$
limit, to the well known ideal gas.

Let us start from the following Hamiltonian $H_0$ describing a
system of $N$ identical particles
\begin{equation}
H_0(\vec{\bfm p})=\sum_{i=1}^N{{\bfm p}_i^2\over2\,m} \
,\label{ham0}
\end{equation}
where $\vec{\bfm p}\equiv ({\bfm p}_1,\,\ldots,\,{\bfm p}_N)$ is the
$3\,N$-vector momenta. In order to be consistent with our model, we
require that the momenta ${\bfm p}_i$ obey the $q$-algebra discussed
in Appendix B. In this case, we can verify that the canonical
distribution $f(\vec{\bfm p})$ obtained from the {\em basic}-entropy
(\ref{qent}) with the mean energy constraint
\begin{equation}
\int\limits_{\mathcal M}H(\vec{\bfm p})\,f(\vec{\bfm
p})\,d_q^{3N}x\,d_q^{3N}p=\langle\epsilon\rangle \ ,
\end{equation}
can be factorized as follows
\begin{equation}
f(\vec {\bfm p})=\prod_{i=1}^Nf({\bfm p}_i) \ .
\end{equation}
Here, $f({\bfm p}_i)$ is the probability distribution function of a
single particle
\begin{equation}
f({\bfm p}_i)={1\over Z_{q,i}}\,{\rm E}_q\left(-\beta_q\,\frac{{\bfm
p}_i^2}{2\,m}\right) \ ,\label{qmb}
\end{equation}
with ${\bfm p}^2_i=p_{x,i}^2+p_{y,i}^2+p_{z,i}^2$ and
\begin{equation}
Z_{q,i}=\int\limits_{\mathcal M}{\rm E}_q\left(-\beta_q\,{{\bfm
p}_i^2\over2\,m}\right)\,d^3_qx_i\,d^3_qp_i=V_{i,q}\,\left(2\,\pi\,m_q\over\beta_q\right)^{3/2}
\ ,\label{i1}
\end{equation}
in units where $\hbar=1$. In the above equation, we have posed
$V_{i,q}=\int_{\mathcal M}d_q^3x_i$ the fractal volume occupied by the
$i$-th particle and $m_q=m\,A_q$ where
\begin{equation}
A_q={1\over\sqrt{\pi}}\int\limits_0\limits^\infty{\rm E}_q(-x^2)\,d_q x
\ ,
\end{equation}
is a constant reducing to unity in the $q\to1$ limit. The canonical
partition function of the whole system is given by
\begin{equation}
Z_{q}=\prod_{i=1}^NZ_{q,i}=V_q^N\left(2\,\pi\,m_q\over\beta_q\right)^{3N/2}
\ .\label{zg}
\end{equation}
We may note that Eq.(\ref{qmb}) can be
interpreted as the $q$-deformed version of the Maxwell-Boltzmann
distribution. It differs formally from the well-known classical
distribution by the mere replacement of the standard exponential
with its $q$-deformed generalization, consistently with the
$q$-algebra underlying the mathematical structure of the theory.

Employing the distribution $f(\vec{\bfm p})$, we can compute the
mean value of any observable associated with the system. In
particular the mean energy is given by
\begin{equation}
\langle\epsilon\rangle=q^{-{3\over2}\,N}\,\left[{3\over2}\,N\right]\,{1\over\beta_q}
\ ,\label{um}
\end{equation}

From Eq.(\ref{um}), we recognize in the $q\to1$ limit the well-known

result $\langle\epsilon\rangle={3\over2}\,N\,T$. We observe that
\begin{equation}
\langle\epsilon\rangle\to +\infty \ \ \ {\rm for} \ q\ll0 \
,\hspace{5mm}{\rm and}\hspace{5mm} \langle\epsilon\rangle\to0 \ \ \
{\rm for} \ q\gg 1 \ ,
\end{equation}
which is a consequence of the dependence of $f(\vec{\bfm p})$ on
$q$. In fact, for smaller and smaller $q$ the tail of the
distribution is enhanced so that particles with high energy give a
larger contribution. In contrast, for larger and larger $q$ the
cut-off inhibits the occupation by the system of the space-phase
cells with high energy and only particles with lower and lower
energy contribute to $\langle\epsilon\rangle$. In particular, for
$q\gg 1$ the distribution $f(\vec{\bfm p})\to\delta(\vec{\bfm0})$ so
that only the fundamental level is populated. Similar arguments can
be applied to justify the expression of the heat capacity given by
\begin{equation}
C_V=q^{-{3\over2}\,N-1}\,\left[{3\over2}\,N\right] \ ,
\end{equation}
which is a constant as in the undeformed classical case, but it is a
monotonically decreasing function reducing to zero for $q\gg 1$.

It is important to outline that the free particle gas with the
Hamiltonian (\ref{ham0}) has the $q$-deformed particle distribution
(\ref{qmb}) only if we require, as a crucial assumption, that the
particle momenta obey the $q$-algebra. On the other hand, if we do
not employ the appropriate $q$-algebra, the same {\it free}
$q$-deformed particle distribution (\ref{qmb}) can be obtained by
assuming the following $N$-body {\it interacting} Hamiltonian
\begin{equation}
H(\vec{\bfm p})=-{1\over \beta_q}\,{\rm Ln}_q\left(\prod_{i=1}^N{\rm
E}_{1/q}\left(\beta_q\,{{\bfm p}_i^2\over2\,m}\right)\right) \
.\label{ham}
\end{equation}
This suggests that the effects of the {\em basic}-deformation of a
{\em free}-ideal gas can be viewed as an effective interaction
described by the Hamiltonian (\ref{ham}). We remark that in the
$q\to1$ limit, the $q$-algebra reduces to the ordinary algebra used
in the Hamiltonian (\ref{ham}). Furthermore, the same Hamiltonian
(\ref{ham}) reduces to the Hamiltonian of a $N$-free particles
system.

It may be important to clarify this point. In several papers
\cite{Scarfone1,Borges,Wang1,Scarfone3}, it has been suggested that,
starting from a deformed exponential derived through physically
and/or mathematically justified arguments, it is possible to
introduce a deformed sum in order to reproduce, in a deformed
fashion, the well-known multiplicative rule of the standard
exponential $\exp(x+y)=\exp(x)\,\exp(y)$. It has been conjectured
(see for instance \cite{Scarfone4,Kaniadakis3}) that such a deformed
sum can be employed, on physical grounds, to take into account the
complex interactions arising among the many-body colliding particles
of a nonlinear medium. This has been discussed, for instance,
explicitly in the Tsallis-entropy framework \cite{Wang2}. In that
case in fact, for the deformed sum of the energy values ${\cal
E}^{\rm A}$ and ${\cal E}^{\rm B}$ belonging to two different
subsystems A and B, it has been assumed that the expression
describes the $q$-sum
\begin{equation}
{\cal E}^{\rm A}\oplus_q{\cal E}^{\rm B}={\cal E}^{\rm A}+{\cal
E}^{\rm B}+{1-q\over\beta}\,{\cal E}^{\rm A}\,{\cal E}^{\rm B} \ ,
\end{equation}
and correspondingly the $q$-Boltzmann factor factorizes according to
\begin{equation}
\exp_q\Big(-\beta\,({\cal E}^{\rm A}\oplus_q{\cal E}^{\rm
B})\Big)=\exp_q\Big(-\beta\,{\cal E}^{\rm
A}\Big)\,\exp_q\Big(-\beta\,{\cal E}^{\rm B}\Big) \ .
\end{equation}
The same situation occurs in the Kaniadakis-entropy framework
\cite{Kaniadakis1} where, by assuming
\begin{equation}
{\cal E}^{\rm A}\stackrel{\kappa}{\oplus}{\cal E}^{\rm B}={\cal
E}^{\rm A}\,\sqrt{1+\Big(\kappa\,\beta\,{\cal E}^{\rm
B}\Big)^2}+{\cal E}^{\rm B}\,\sqrt{1+\Big(\kappa\,\beta\,{\cal
E}^{\rm A}\Big)^2} \ ,
\end{equation}
for the energy levels, it has been shown that the $\kappa$-Boltzmann
factor factorizes in
\begin{equation}
\exp_{\{\kappa\}}\Big(-\beta\,({\cal E}^{\rm
A}\stackrel{\kappa}{\oplus}{\cal E}^{\rm
B})\Big)=\exp_{\{\kappa\}}\Big(-\beta\,{\cal E}^{\rm
A}\Big)\,\exp_{\{\kappa\}}\Big(-\beta\,{\cal E}^{\rm B}\Big) \ .
\end{equation}
Of course, all of this can be also reproduced within the formalism
employed in the present work. In fact, we can verify that the
following deformed sum
\begin{equation}
{\cal E}^{\rm A}\oplus{\cal E}^{\rm B}=-{1\over\beta}\,{\rm
Ln}_q\Big({\rm E}_{1/q}\Big(\beta\,{\cal E}^{\rm A}\Big)\,{\rm
E}_{1/q}\Big(\beta\,{\cal E}^{\rm A}\Big)\Big) \ ,\label{qsum}
\end{equation}
which reduces to the ordinary sum in the $q\to1$ limit, fulfils the
factorization rule
\begin{equation}
{\rm E}_q\Big(-\beta\,({\cal E}^{\rm A}\oplus{\cal E}^{\rm
B})\Big)={\rm E}_q\Big(-\beta\,{\cal E}^{\rm A}\Big)\,{\rm
E}_q\Big(-\beta\,{\cal E}^{\rm B}\Big) \ ,\label{eq}
\end{equation}
for the {\em basic}-Boltzmann factor. We easily recognize in the
definition (\ref{qsum}) the origin of the Hamiltonian (\ref{ham}).
Although it is beyond the scope of the present paper, it is also
possible to show that the {\em basic}-sum (\ref{qsum}) obeys all the
axiomatic properties where a well defined sum must satisfy
associativity, commutativity, existence and uniqueness of the
inverse element and of the identity element. However, by applying
the appropriate $q$-algebra introduced starting from the
$q$-binomial expansion , the deformed sum $x\oplus y$ is replaced,
in a natural way, by the ordinary sum $x+y$. This aspect makes the
basic thermostatistics formalism very interesting, because of the
structure of the considered deformation implies a close and
consistent realization in the well known mathematical framework of
$q$-calculus.


\sect{Conclusion}
\renewcommand{\theequation}{\arabic{section}.\arabic{equation}}

In this paper we have studied a possible generalization of the
thermostatistics theory of a classical system based on the
$q$-deformed algebra. Starting from the definition of the {\em
basic}-exponential, we have introduced a generalized entropic
function and derived, by means of a consistent $q$-variational
principle, a deformed probability distribution function which
differs from the standard Gibbs distribution by the replacement of
the ordinary exponential function with its generalization furnished
by the {\em basic}-exponential. We have performed a preliminary
investigation of some fundamental thermodynamic relations which are
preserved consistently with the formalism of the $q$-calculus.

On physical grounds, it has been demonstrated that the distribution
arising in this model exhibits a cut-off in the energy
spectrum which is generally expected in those systems whose
underlying dynamics is governed by long-range interactions. Such a
feature has been observed also in other distributions proposed in
the literature, obtained from generalized versions of the
Boltzamnn-Gibbs entropy. What is different here, is the asymptotic
behavior of the distribution derived in this paper which does not
match with the power-law behavior typically shown by the other
generalized distributions.

We have studied, within the present formalism, a $N$-body system of
interacting particles described by the Hamiltonian (\ref{ham}),
whose interaction vanishes in the $q\to1$ limit. By construction,
the canonical distribution function of this system is formally
equivalent to the one derived starting from the Hamiltonian
(\ref{ham0}), describing a system of non-interacting particles,
where the momenta obeying the $q$-algebra originates from the
$q$-binomial expansion (\ref{qbin}). In this sense, the toy model
described by the Hamiltonian (\ref{ham}) can be considered as the
$q$-analog of the ideal gas.

Being Jackson $q$-derivative and $q$-integral the natural tools for
describing discrete-scale invariance \cite{erzan1,erzan2}, we expect
that this study can be a very useful starting point on the {\em
basic}-thermostatistic theory which can be strictly related to
critical phenomena (such as growth processes, rupture, earthquake,
financial crashes) with the existence of log-periodic oscillations
deriving from a partial breakdown of the continuous scale-invariance
symmetry into a discrete-scale-invariance symmetry, as occurs for
instance in hierarchical lattice \cite{sornette}. In fact, as
mentioned in the Subsections 3.1 and 3.2, our study on the {\em
basic}-thermostatistics incorporates implicitly a self-similarity in
the parameter space, labeling the phase space of the system, and,
consequently, a fractal structure in the energy spectrum emerges. In
this context, it is remarkable to observe that, for example, the
specific heat corresponding to systems with deterministic fractal
energy is known to present log-periodic oscillations as a function
of the temperature around a mean value given by a characteristic
dimension of the energy spectrum \cite{coronado,tsallis,anteneodo}.
A detailed study of these critical phenomena in our formalism lies
out the scope of this paper and will be the matter of future
investigations.

\app \sect{} As known, a possible way to obtain the stationary
distribution of a system governed by a given entropy, under a set of
suitable physically constraints, follows by means of the variational
calculus on the constrained entropic form. In the case under
investigation, accounting the $q$-algebra underling our
formalism, we have to deal with the following problem
\begin{equation}
\delta\,{\cal F}(p)=0 \ ,\label{b1}
\end{equation}
where, according to Eq.(\ref{var2})
\begin{equation}
{\cal F}(p)=-\int\limits_{\cal M}p(\lambda)\,\Big[{\rm
Ln}_q(p(\lambda))+\sum_j\,\mu_j^\ast\,\phi_j(\lambda)\Big]\,d_q\lambda
\ .\label{b2}
\end{equation}
Without loss of generality, we pose $\mu_0^\ast=q^{-1}(1+\mu_0)-1$
and $\mu_i^\ast=q^{-1}\,\mu_i$ for $i=1,\,\ldots M$, where $\mu_j$ are
the Lagrange multipliers of to the $M+1$ constraints (\ref{constr}) introduced in the Section 3.\\
By using the ansatz (\ref{ans}) in ${\cal F}(p)\equiv \widetilde {\mathcal F}(p(f))$, Eq.(\ref{b1}) can
be computed as follows
\begin{eqnarray}
\nonumber&&\hspace{-25mm}\delta\,\widetilde{\cal
F}(f)=\lim_{t\to0}\Bigg\{\int\limits_{\cal
M}\Big[f(\lambda)+t\,h(\lambda)-\sum_j\mu_j^\ast\,
\phi_j(\lambda)\Big]\,{\rm
E}_q\Big(-f(\lambda)-t\,h(\lambda)\Big)\,d_q\lambda\\
\nonumber &&-\int\limits_{\cal M}\Big[f(\lambda)-\sum_j\mu_j^\ast\,
\phi_j(\lambda)\Big]\,{\rm
E}_q\Big(-f(\lambda)\Big)\,d_q\lambda\Bigg\}\\
\nonumber &&\hspace{-25mm}=\frac{d}{d\,t}\int\limits_{\cal
M}\Big[f(\lambda)+t\,h(\lambda)-\sum_j\mu_j^\ast\,
\phi_j(\lambda)\Big]\,{\rm
E}_q\Big(-f(\lambda)-t\,h(\lambda)\Big)\,d_q\lambda\Bigg|_{t=0}\\
\nonumber&&\hspace{-25mm}=\int\limits_{\cal
M}\Bigg[h(\lambda)+\Big(f(\lambda)+t\,h(\lambda)-\sum_j\mu_j^\ast\,
\phi_j(\lambda)\Big)\frac{\frac{d}{d\,t}{\rm
E}_{1/q}\Big(-t\,h(\lambda)\Big)}{{\rm
E}_{1/q}\Big(-t\,h(\lambda)\Big)}\Bigg]_{t=0}\,{\rm
E}_q\Big(-f(\lambda)\Big)\,d_q\lambda \ ,\\ \label{b5}
\end{eqnarray}
where ${\rm E}_q(-f-t\,h)={\rm E}_q(-f)\,{\rm E}_{1/q}(-t\,h)$
according to Eq.(\ref{com1}).\\
By taking into account the definition (\ref{qexp}) we have
\begin{equation}
\frac{d}{dt}{\rm
E}_{1/q}\Big(-t\,h(\lambda)\Big)=\sum_{n=1}\frac{n}{[n]!}\Big(-h(\lambda)\Big)\,t^{n-1}\Bigg|_{t=0}=-h(\lambda)
\ ,
\end{equation}
so that from Eq.(\ref{b5}) we obtain
\begin{eqnarray}
\delta\,\widetilde{\cal F}(f)=\int\limits_{\cal
M}h(\lambda)\Big(1-f(\lambda)+\sum_j\mu_j^\ast\,
\phi_j(\lambda)\Big)\,{\rm E}_q\Big(-f(\lambda)\Big)\,d_q\lambda=0 \
,\label{b4}
\end{eqnarray}
Accounting for the arbitrariness of the function $h(\lambda)$ this
last equation implies
\begin{equation}
f(\lambda)=1+\sum_j\mu_j^\ast\,\phi_j(\lambda) \ ,
\end{equation}
in accordance with Eq. (\ref{distr}) given in the text.

\sect{} We present a generalization of the algebra (\ref{qbin}) to a
trinomial with the purpose of extending the factorization formula of
the {\em
basic}-exponential.\\
First, let us briefly review the derivation of Eq.(\ref{com1}).\\
This can be shown easily by considering the Cauchy product among the
series (\ref{qexp}) and its analogue for $q\to1/q$. We obtain
\begin{eqnarray}
\nonumber \hspace{-20mm}{\rm E}_q(x)\,{\rm
E}_{1/q}(y)&=&1+\Bigg({x\over[1]!}+{y\over[1]!}\Bigg)+\Bigg({x^2\over[2]!}+{x\,y\over[1]![1]!}+
{y^2\over[2]!}\Bigg)+\ldots+\\
&+&\Bigg({x^n\over[n]!}+{x^{n-1}\,y\over[n-1]!\,[1]!}+
{q\,x^{n-2}\,y^2\over[n-2]!\,[2]!}+\ldots+{q^{n\,(n-1)/2}\,y^n\over[n]!}\Bigg)+\ldots
\ , \label{aaa1}
\end{eqnarray}
which, by means of Eq.(\ref{qbin}), can be rewritten in the form
\begin{equation}
{\rm E}_q(x)\,{\rm
E}_{1/q}(y)=1+{(x+y)^{(1)}\over[1]!}+{(x+y)^{(2)}\over[2]!}+\ldots+{(x+y)^{(n)}\over[n]!}+\ldots
\ ,
\end{equation}
and coincides with the definition of ${\rm E}_q(x+y)$ given in Eq.(\ref{expq1}).\\
To generalize this result, we consider the following $q$-binomial
expansions
\begin{eqnarray}
\nonumber &&(x+z)^{(0)}=1 \ ,\\
\nonumber
&&(x+z)^{(1)}=x+z \ ,\\
\nonumber &&(x+z)^{(2)}=x^2+[2]\,x\,z+q\,z^2 \ ,\\
&&(x+z)^{(3)}=x^3+[3]\,x^2\,z+q\,[3]\,x\,z^2+q^3\,z^3 \ ,\label{aa4}
\end{eqnarray}
and so on. By redefining $x\to x+y$ and consequently
$x^n\to(x+y)^{(n)}$, we obtain
\begin{eqnarray}
\nonumber ((x+y)+z)^{(0)}&=&1 \ ,\\
\nonumber
((x+y)+z)^{(1)}&=&x+y+z \ ,\\
\nonumber ((x+y)+z)^{(2)}&=&(x+y)^{(2)}+[2]\,(x+y)^{(1)}\,z+q\,z^2\\
\nonumber
&=&x^2+[2]\,x\,y+q\,y^2+[2]\,x\,z+[2]\,y\,z+q\,z^2 \ ,\\
\nonumber ((x+y)+z)^{(3)}&=&(x+y)^{(3)}+[3]\,(x+y)^{(2)}\,z+q\,[3]\,(x+y)^{(1)}\,z^2+q^3\,z^3\\
\nonumber
&=&x^3+[3]\,x^2\,y+q\,[3]\,x\,y^2+q^3\,y^3+[3]\,x^2\,z\\
& &+[2]\,x\,y\,z+q\,y^2\,z +q\,[3]\,x\,z^2+q\,[3]\,y\,z^2+q^3\,z^3
,\label{aa3}
\end{eqnarray}
which implies the following factorization rule for the {\em
basic}-exponential
\begin{equation}
{\rm E}_q(x+y+z)={\rm E}_q(x+y)\,{\rm E}_{1/q}(z)={\rm E}_q(x)\,{\rm
E}_{1/q}(y)\,{\rm E}_{1/q}(z) \ .\label{aa2}
\end{equation}
On the other hand, starting from the $q$-binomial expansion
(\ref{aa4}) and by posing $z^n\to (y+z)^{(n)}_{1/q}$ \footnote{We
have introduced the index $1/q$ to indicate the replacement
$q\to1/q$ in the expansion of the $q$-binomial given in Eq.
(\ref{qbin})}we form the $q$-trinomial expansion
$(x+(y+z)_{1/q})^{(n)}$ which implies the following decomposition
\begin{equation}
{\rm E}_q(x+y+z)={\rm E}_q(x)\,{\rm E}_{1/q}(y+z)={\rm E}_q(x)\,{\rm
E}_{1/q}(y)\,{\rm E}_{q}(z) \ .
\end{equation}
Other possible factorization rules can be realized through the
introduction of suitable $q$-deformed algebras, as can be seen by
inspection. Extension to an arbitrary polynomial can be also easily
obtained.

In the following, let us show the use of the above algebra in the
derivation of the distributions (\ref{mm}), (\ref{can}) and
(\ref{gcan}). Starting from the equality
\begin{equation}
x+y=0 \ ,\label{aa1}
\end{equation}
and employing the $q$-algebra (\ref{aa4}) we can construct the
$q$-binomial $(x+y)^{(n)}=0$ which of course vanishes for all $n>0$.
Consequently, according to the definition (\ref{qexp}), by dividing Eq.(\ref{aa1})
by $[n]!$ and summing up over
$n=0,\,\ldots,\,\infty$, we obtain
\begin{equation}
{\rm E}_q(x+y)={\rm E}_q(x)\,{\rm E}_{1/q}(y)=1 \ .
\end{equation}
In particular, by applying this result to Eq.(\ref{qme2}) which has
the form (\ref{aa1}), with $x={\rm Ln}_q\,p(\epsilon)$ and
$y=q^{-1}(1+\alpha)$, we obtain
\begin{equation}
{\rm E}_q\Big({\rm Ln}_q\,p(\epsilon)\Big)\,{\rm
E}_{1/q}\Big(q^{-1}\,(1+\alpha)\Big)=1 \ ,
\end{equation}
so that
\begin{equation}
p(\epsilon)={\rm E}_q\Big(-q^{-1}\,(1+\alpha)\Big) \ ,
\end{equation}
and according to the property (\ref{inv}) we obtain Eq.(\ref{mm}) given in Section III-A.\\
In the same manner, from the equality
\begin{equation}
{\rm Ln}_q\,p(\epsilon)+q^{-1}\,(1+\alpha+\beta\,\epsilon)=0 \ ,
\end{equation}
given in Eq.(\ref{eq1}), employing the $q$-algebra (\ref{aa3}) with
$x={\rm Ln}_q\,p(\epsilon)$, $y=q^{-1}\,(1+\alpha)$ and
$z=q^{-1}\,\beta\,\epsilon$, it follows that
\begin{equation}
{\rm E}_q\Big({\rm Ln}_q\,p(\epsilon)\Big)\,{\rm
E}_{1/q}\Big(q^{-1}\,(1+\alpha)\Big)\,{\rm
E}_{1/q}\Big(q^{-1}\,\beta\,\epsilon\Big)=1 \ ,
\end{equation}
or equivalently
\begin{equation}
p_i={{\rm E}_q\Big(-q^{-1}\,(1+\alpha)\Big)\,{\rm
E}_q\Big(-q^{-1}\,\beta\,\epsilon\Big)} \ ,
\end{equation}
and by using again the property (\ref{inv}) it can be written in the
form (\ref{can}) given in Section III-B. Similar arguments can be
applies to obtain the distribution (\ref{gcan}) by employing the
appropriate $q$-algebra with $x={\rm Ln}_q\,p(\epsilon)$,
$y=q^{-1}\,(1+\alpha)$, $z=q^{-1}\,\beta\,\epsilon$ and
$u=q^{-1}\,\mu\,\beta\,N$.

\vspace{10mm} \noindent{\bf Acknowledgments}\\

\noindent Authors wish to thank an anonymous referee to drive our
attention to the $q$-integral calculus, its relation to fractal sets
and useful suggestions. It is also a pleasure to thank Prof. P.
Quarati for fruitful comments and suggestions.



\section*{References}



\begin{thebibliography}{99}




\bibitem{Beck}
Beck C and Schl\"{o}gl F 1993 {\em Thermodynamics of chaotic system}
(Cambridge University Press)




\bibitem{Hilborn}
Hilborn R 2001 {\em Chaos and Nonlinear Dynamics: An Introduction
for Scientists and Engineers} (Oxford University Press).




\bibitem{Abe01}
Abe S 2001 {\em Nonextensive Statistical Mechanics and its
Applications} ed Y Okamoto (Springer 2001)




\bibitem{GellMann}
Gell-Mann M and Tsallis C 2004 {\em Nonextensive Entropy:
Interdisciplinary Applications} (Oxford University Press)










\bibitem{Tsallis1}
Tsallis C 1988 J. Stat. Phys. {\bf52} 479; Curado E M and Tsallis C
1991 J. Phys. A {\bf 24} L69; Tsallis C, Mendes R S and Plastino A R
1998 Physica A {\bf 261} 534




\bibitem{Abe} Abe S 1997 Phys. Lett. A {\bf224} 326




\bibitem{Kaniadakis1} Kaniadakis G 2002 Phys. Rev. E {\bf66}
056125




\bibitem{Scarfone1} Kaniadakis G, Lissia M and Scarfone A M 2004
Physica A {\bf 340} 41; 2005 Phys. Rev. E {\bf71} 046128




\bibitem{Scarfone03} Scarfone A M and Wada T 2005 Phys. Rev. E
{\bf72} 026123 (2005)




\bibitem{Scarfone2} Scarfone A M 2006 Phys. Lett. A {\bf355} 404




\bibitem{Tsallis2} See http://tsallis.cat.cbpf.br/biblio.htm for an updated bibliography on
the subject.




\bibitem{wil} Wilczek F 1990 {\em Fractional Statistics and Anyon Superconductivity} (World
Scientific, Singapore) and references therein.




\bibitem{bie} Biedenharn L 1989 J. Phys. A: Math. Gen. {\bf 22} L873




\bibitem{mac} Macfarlane A 1989 J. Phys. A: Math. Gen. {\bf 22} 4581




\bibitem{heine}
Heine E 1846 J. reine angew. Math. {\bf 32} 210; 1847 {\bf 34} 285;
1878 {\em Handbuch der Kugelfunctionen, Theorie und Anwendungen}
Vol. 1 (Reimer, Berlin)




\bibitem{jack}
Jackson F H 1909 Am. J. Math. {\bf 38} 26; 1909 Mess. Math. {\bf 38}
57




\bibitem{gasper}
Gasper G and Rahman M 1990 {\em Basic hypergeometric series},
Encyclopedia of mathematics and its applications (Cambridge
Univeristy Press)





\bibitem{Exton}
Exton H 1983 {\em q-Hypergeometric functions and applications}
(Chichester: Ellis Horwood)



\bibitem{cele1} Celeghini E {\em et al.} 1995 Ann. Phys. {\bf 241} 50




\bibitem{fink} Finkelstein R J 1998 Int. J. Mod. Phys. A {\bf 13} 1795




\bibitem{Lavagno1}
Lavagno A and Swamy N P 2000 Phys. Rev. E {\bf 61} 1218; 2002 Phys.
Rev. E {\bf65} 036101



\bibitem{erzan1} Erzan A and Eckmann J -P 1997 Phys. Rev. Lett.
{\bf78} 3245



\bibitem{erzan2} Erzan A 1997 Phys. Lett. A {\bf 225} 235



\bibitem{marmo1} Bimonte G, Esposito C, Marmo G and Stornaiolo C 2003 Phys. Lett. A
{\bf 318} 313




\bibitem{noi1}
Lavagno A, Scarfone A M and Swamy N P 2005 Rep. Math. Phys. {\bf55}
423; 2006 Eur. Phys. J. C {\bf 47} 253.




\bibitem{noi3} Lavagno A, Scarfone A M and Swamy N P 2006 Eur.
Phys. J. B {\bf50}, 351




\bibitem{Spitzer} Spitzer Jr. L 1940 MNRAS {\bf100} 396




\bibitem{boer}
de Boer J {\it et al.} 1994 Phys. Rev. Lett. {\bf 73} 906




\bibitem{Plastino2} Plastino A R and Plastino A 1994 Phys. Lett. A
{\bf193} 140




\bibitem{lee}
Lee M -H and Kim J K 1996 Phys. Rev. D {\bf 54} 3904




\bibitem{Lavagno0} Lavagno A, Kaniadakis G, Rego-Monteiro M,
Quarati P and Tsallis C 1998 Astrophys. Lett. \& Comm. {\bf35} 449




\bibitem{kami}
Kaminski A and Glazman L I 2001 Phys. Rev. Lett. {\bf 86} 2400




\bibitem{vavro}
Vavro J {\it et al.} 2003 Phys. Rev. Lett. {\bf 90} 065503




\bibitem{naza}
Nazaretski E {\it et al.} 2005 Phys. Rev. B {\bf 71} 144201




\bibitem{loeb}
Loeb A and Zaldarriaga M 2005 Phys. Rev. D {\bf 71} 103520




\bibitem{Wess} Cerchiai B L, Hinterding R, Madore J and Wess J 1999
Eur. J. Phys C {\bf8} 547




\bibitem{Kulish} Kulish P P and  Damaskinsky E V 1990 J.
Phys. A: Math. Gen. {\bf 23} L415




\bibitem{Yang} Yang Y and Yu Z 1994 Mod. Phys. Lett. A {\bf9} 3367





\bibitem{Ubriaco} Ubriaco M R 1992 J. Phys. A: Math. Gen. {\bf25}
169




\bibitem{Klimek} Klimek M 1993 J. Phys. A: Math. Gen. {\bf26} 955




\bibitem{Charles} Nelson A C and Gartley M G 1996 J. Phys. A: Math.
Gen. {\bf29} 8099




\bibitem{Ward} Ward M 1936 Am. J. Math. {\bf58} 255




\bibitem{Sharma} Sharma B D and Mittal D P 1975 J. Math.
Sci. {\bf10} 28




\bibitem{Renyi} R\'{e}nyi A 1970 {\em Probability theory}
(Amsterdam: North-Holland Publ. Company)



\bibitem{Callen} Callen H B 1985 {\em Thermodynamics and an
Introduction to Thermostatistics} (New York: Wiley)




\bibitem{Borges} Borges E P 2004 Physica A {\bf340} 95




\bibitem{Wang1} Nivanen L, Le M\'{e}haut\'{e} A and Wang Q A 2003
Rep. Math. Phys. {\bf52} 437




\bibitem{Scarfone3} Kaniadakis G and Scarfone A M 2002 Physica A
{\bf305} 69




\bibitem{Scarfone4} Kaniadakis G, Quarati P
and Scarfone A M 2002 Physica A {\bf305} 76




\bibitem{Kaniadakis3} Bir\'{o} T S and Kaniadakis G 2006 Eur. J. Phys.
B {\bf50} 3




\bibitem{Wang2} Wang  Q A 2002 Eur. Phys. J. B {\bf26} 357



\bibitem{sornette}
Gluzman S and Sornette D 2002 Phys. Rev. E {\bf 65} 036142



\bibitem{coronado}
Coronado A V and Carpena 2006 Phys. Rev. E {\bf 73} 016124



\bibitem{tsallis}
Tsallis C {\it et al.} 1997 Phys. Rev. E {\bf 56} R4922



\bibitem{anteneodo}
Vallejos R O and Anteneodo C Phys. Rev. E {\bf 58} 4134



\end{thebibliography}
\end{document}